\documentclass{ws-ijqi}

\begin{document}

\markboth{G. Rigolin}
{Thermal Entanglement in the Two-Qubit Heisenberg XYZ Model}
 
\title{THERMAL ENTANGLEMENT IN THE TWO-QUBIT HEISENBERG XYZ MODEL}

\author{GUSTAVO RIGOLIN}

\address{Departamento de Raios C\'osmicos e Cronologia, Instituto de F\'{\i}sica Gleb Wataghin, Universidade Estadual de Campinas, C.P. 6165, cep 13084-971, Campinas, S\~ao Paulo, Brazil \\
rigolin@ifi.unicamp.br}

\maketitle

\begin{abstract}
We study the entanglement of a two-qubit one dimensional XYZ Heisenberg chain in thermal equilibrium at temperature T. We obtain an analytical expression for the concurrence of this system in terms of the parameters of the Hamiltonian and T. We show that depending on the relation among the coupling constants it is possible to increase the amount of entanglement of the system increasing its anisotropy. We also show numerically that for all sets of the coupling constants entanglement is a monotonically decreasing function of the temperature T, proving that we must have at least an external magnetic field in the z-direction to obtain a behavior where entanglement increases with T.        
\end{abstract}


\section{Introduction}

Since the beginning of Quantum Mechanics (QM) it was recognized that the superposition principle implied novel and counter-intuitive experimental results \cite{epr,schroedinger,bell}. Erwin Schr\"odinger used the word entanglement to succinctly express one of the most striking non-local features of that new theory. However, for a long time since the discovery of entanglement, nobody expected that it would have some practical application and the majority of the physicists thought that it would be restricted to the conceptual discussions of QM. Nevertheless, the last decade of the twentieth century revealed incredible practical applications for the entanglement. From new quantum algorithms that outperform their classical counterparts \cite{shor,grover} to quantum communication \cite{superbennett,telebennett,zeilinger} we see the usefulness of quantum entanglement.

Once accepted that entanglement is a resource we can manipulate to do useful tasks, it became a necessity to quantify it. It turned out that this quantification is not an easy task and up to now it is one of the most challenging open questions in Quantum Information Theory \cite{formation,destilation,relative}. Fortunately, for bipartite systems of two levels, i. e.,  a pair of qubits, there exists an analytical expression to quantify its amount of entanglement \cite{wootters}. It is called Entanglement of Formation ($E_{F}$) \cite{formation}. Given the density matrix $\rho$ that describes our pair of qubits, $E_{F}$ is the average entanglement of the pure states of the decomposition of $\rho$, minimized over all possible decompositions:
\begin{equation}
E_{F}(\rho) = \mbox{min}\sum_{i}p_{i}E(\psi_{i}),
\end{equation}
where $\sum_{i}p_{i} = 1$, $0 < p_{i} \leq 1$, and $\rho = \sum_{i}p_{i}\left| \psi_{i}\right>\left< \psi_{i}\right|$. Here $E(\psi)$ is the von Neumann entropy of either of the two qubits \cite{bennett}. For a pair of qubits Wootters \textit{et al} \cite{wootters} have shown that $E_{F}$ is a monotonically increasing function of the concurrence $C$, which one can prove to be an entanglement monotone. When $C=1$ we have maximally entangled states and when $C=0$ we do not have entanglement. Since $C$ is mathematically simpler to deal with than $E_{F}$ and that knowing it we can automatically get $E_{F}$, we concentrate our efforts calculating $C$ to study the amount of entanglement between two qubits. The concurrence between them is \cite{wootters}:
\begin{equation}
C = \mbox{max} 
\{ \lambda_{1} - \lambda_{2} - \lambda_{3} - \lambda_{4}, 0 \},
\label{concurrence}
\end{equation}
where $\lambda_{1},  \lambda_{2}, \lambda_{3}$, and $\lambda_{4}$
are the square roots of the eigenvalues, in decreasing order, of the matrix $R =\rho\tilde{\rho}$. Here $\tilde{\rho}$ is the spin flipped matrix given by
\begin{equation}
\tilde{\rho} = \left(\sigma_{y} \otimes \sigma_{y}\right) \rho^{*} \left(\sigma_{y} \otimes \sigma_{y}\right).
\end{equation} 
The symbol $\rho^{*}$ means complex conjugation of the matrix $\rho$ in the standard basis $\left\{ \left| 00 \right>, \left| 01 \right>, \left| 10 \right>, \left| 11 \right> \right\}$. 

Recently, considerable efforts have been devoted in the study of Heisenberg spin systems \cite{nielsen,vedral,kamta,wang,chen,li} concerning their amount of entanglement for some temperature T. These systems are relatively simple and can describe real solid state systems \cite{hammar}, which can possibly be used to generate entangled qubits \cite{lea1,lea2,lea3} or used in the construction of quantum gates \cite{imamoglu,raussendorf}, the building blocks of any quantum computer.

In this article we deal with 1D Heisenberg open chains with no external magnetic field and only nearest neighbor interactions. This system is often called XYZ model and the Hamiltonian that describes the system is:
\begin{equation}
H = \sum_{i=1}^{N-1}\left( \frac{J_{x}}{4}\sigma_{x}^{i}\sigma_{x}^{i+1} + \frac{J_{y}}{4}\sigma_{y}^{i}\sigma_{y}^{i+1}+  \frac{J_{z}}{4}\sigma_{z}^{i}\sigma_{z}^{i+1} \right),
\label{xyz}
\end{equation}        
where we restrict to chains of only two spins ($N=2$), $J_{x}, J_{y}, J_{z}$ are the coupling constants, $\sigma_{x}, \sigma_{y}, \sigma_{z}$ are the Pauli matrices, and $\hbar = 1$. 
 
Here we present an analytical formula for the concurrence and a detailed study for the general XYZ model. We show that there exist regions for the XY and XYZ models where the concurrence \textit{increases} monotonically as we increase the anisotropy of the system. This behavior is a new feature of thermal entanglement in chains with no external magnetic field, since Wang \cite{wang} and Kamta \textit{et al} \cite{kamta} only studied systems where the concurrence decreases with anisotropy. In these regions the anisotropy also \textit{increases} the critical temperature $T_{c}$ beyond which the concurrence is zero. We review the results obtained by Refs.~\refcite{kamta}-\refcite{chen} for the XY model ($J_{z} = 0$) and by Ref.~\refcite{vedral} for the XXX model ($J_{x}=J_{y}=J_{z}$), which are obtained as particular cases of our general solution. We also numerically show that contrary to the cases where we have an external magnetic field \cite{vedral,kamta,wang,chen}, there does not exist any set of the coupling constants which allows an increase of the concurrence as we increase the temperature T of the system. 

\section{XYZ Thermal State: An Overview}

In order to study the XYZ model we rewrite the Hamiltonian (\ref{xyz}) in the following form:
\begin{equation}
H = \frac{J_{z}}{4}\sigma_{z}^{1}\sigma_{z}^{2} + \frac{ \Sigma + \Delta}{8}\sigma_{x}^{1}\sigma_{x}^{2}+  \frac{\Sigma - \Delta}{8}\sigma_{y}^{1}\sigma_{y}^{2},
\label{xyz2}
\end{equation}
where $\Delta = J_{x} - J_{y}$ and $\Sigma = J_{x} + J_{y}$. The four eigenvectors of the Hamiltonian are the four Bell states (maximally entangled states): $H\left| \Phi^{\pm} \right> = \lambda_{\Phi^{\pm}}\left| \Phi^{\pm} \right>$ and $H\left| \Psi^{\pm} \right> = \lambda_{\Psi^{\pm}}\left| \Psi^{\pm} \right>$, where $\left| \Phi^{\pm} \right> = \left(1/\sqrt{2}\right)\left(\left| 00 \right> \pm \left| 11 \right>\right)$, $\left| \Psi^{\pm} \right> = \left(1/\sqrt{2}\right)\left(\left| 01 \right> \pm \left| 10 \right>\right)$, $\lambda_{\Phi^{\pm}} = \left(J_{z} \pm \Delta \right)/4$, and $\lambda_{\Psi^{\pm}} = \left(- J_{z} \pm \Sigma \right)/4$. 

We use the parameter $\delta = \Delta/\Sigma$ to measure the anisotropy of the system \cite{kamta}. When $\delta = 0$ and $J_{z} = 0$ we have the isotropic XY model and when $\delta = \pm 1$ and $J_{z} = 0$ we have the Ising model. 

The density matrix which describes a system in equilibrium with a thermal reservoir at temperature T (canonical ensemble) is $\rho = \exp{\left( -H/kT  \right)}/Z$, where $Z = \mbox{Tr}\left\{ \exp{\left( -H/kT  \right)} \right\}$ is the partition function and $k$ is Boltzmann's constant. Hamiltonian (\ref{xyz2}) gives the following thermal state written in the standard basis:
%
\begin{equation}
\rho  =  \frac{1}{Z} 
\left( 
\begin{array}{cccc}
e^{-\alpha}\cosh(\beta) & 0 & 0 & -e^{-\alpha}\sinh(\beta) \\
0 & e^{\alpha}\cosh(\gamma) & - e^{\alpha}\sinh(\gamma)  & 0 \\ 
0 & -e^{\alpha}\sinh(\gamma) &  e^{\alpha}\cosh(\gamma)  & 0 \\
-e^{-\alpha}\sinh(\beta) & 0 & 0 & e^{-\alpha}\cosh(\beta) \\
\end{array}
\right), \label{rho}
\end{equation} 
%
where $\alpha = J_{z}/(4kT)$, $\beta = \Delta/(4kT)$, $\gamma = \Sigma/(4kT)$, and $Z = 2\left( \exp{(-\alpha)}\cosh(\beta) + \exp{(\alpha)}\cosh(\gamma) \right)$.

The square roots of the four eigenvalues of the matrix $R =\rho\tilde{\rho}$, which enable us to obtain the concurrence, are:
\begin{eqnarray}
\lambda_{I}^{\pm} & = & \frac{e^{-\alpha}}{Z}\left( \cosh(\beta) \pm \sinh(\beta)  \right), \\
\lambda_{II}^{\pm} & = & \frac{e^{\alpha}}{Z}\left( \cosh(\gamma) \pm \sinh(\gamma)  \right). 
\label{squareeigen}  
\end{eqnarray}
It is not a trivial task to put in decreasing order $\lambda_{I}^{\pm}$ and $\lambda_{II}^{\pm}$, since we need the values of $\alpha, \beta$, and $\gamma$ to correctly arrange them. However, the concurrence of $\rho$, Eq.~(\ref{concurrence}), can be written in the following somewhat simple analytical form:
\begin{equation} 
\label{geralxyz}
C = 
\left\{
\begin{array}{cc}
\mbox{max}\left\{ 0,\, C_{1} \right\}, &  \mbox{if} \; 2\alpha >  |\beta|- |\gamma|, \\
\mbox{max}\left\{ 0,\, C_{2} \right\}, & \mbox{if} \; 2\alpha \le  |\beta|- |\gamma|, 
\end{array}
\right.
\end{equation}
where
\begin{eqnarray}
C_{1} & = & \frac{e^{\alpha}\sinh(|\gamma|)-e^{-\alpha}\cosh(\beta)}{e^{\alpha}\cosh(\gamma)+e^{-\alpha}\cosh(\beta)}, \\
C_{2} & = & \frac{e^{-\alpha}\sinh(|\beta|)-e^{\alpha}\cosh(\gamma)}{e^{\alpha}\cosh(\gamma)+e^{-\alpha}\cosh(\beta)}.
\end{eqnarray}
Before we study in detail Eq.~(\ref{geralxyz}) for arbitrary values of $J_{x}, J_{y}$, and $J_{z}$ we will analyze the concurrence for some particular interesting cases. 

\section{Ising Model}
In the Ising model $J_{x}=J_{y}=0$ \cite{wang}. This implies that $\beta = \gamma =0$. Substituting in  Eq.~(\ref{geralxyz}) we obtain:
\begin{equation}
C = \mbox{max}\left\{ 0, \frac{-e^{-|\alpha|}}{e^{|\alpha|}+e^{-|\alpha|}} \right\} = 0.  
\end{equation} 
We can understand why the thermal Ising system is not entangled for any T looking at the density matrix (\ref{rho}). When  $\beta = \gamma =0$ it is diagonal in the standard basis implying no quantum correlations. This result is not surprise because despite of having four maximally entangled states as eigenvectors, $\left| \Phi^{\pm} \right>$ and $\left| \Psi^{\pm} \right>$ are degenerated, which implies that the Ising thermal state has no entanglement. 

\section{XY Model}

When $J_{z} = 0$ we deal with the XY model. It is called isotropic XY model when $J_{x} = J_{y} = J$ and anisotropic XY model when $J_{x} \ne J_{y}$. We study separately the two cases. 

\subsection{Isotropic Case}

In the isotropic XY model $\alpha = \beta = 0$ and $\gamma = J/(2kT)$. Using these values in Eq.~(\ref{geralxyz}) we get:
\begin{equation}
C = \mbox{max}\left\{ 0, \frac{\sinh\left(\frac{|J|}{2kT}\right) - 1}{\cosh\left(\frac{J}{2kT}\right) + 1} \right\}.
\label{xyiso}
\end{equation}
Analyzing Eq.~(\ref{xyiso}) we see that for very low temperatures we have concurrence close to $1$ and that it decreases monotonically with the temperature until a critical temperature $T_{c}$ \cite{wang}, which is the solution of $\sinh(|J|/(2kT_{c})) = 1$. We also see that systems with strong coupling (higher $J$) have a greater concurrence for a given $T$ than those systems with weak coupling and that the concurrence is the same whether the system is ferromagnetic ($J<0$) or antiferromagnetic ($J>0$).  

\subsection{Anisotropic Case}
\begin{figure}[ht]
\centerline{\psfig{file=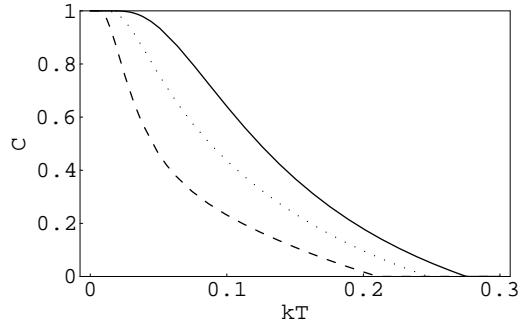, width=7cm}}
\vspace*{8pt}
\caption{ The dependence of the concurrence $C$ for the XY model with the absolute temperature $kT$. The solid line represents $\delta = 0.3$, the dotted line shows $\delta = 0.6$, and the dashed line is for  $\delta = 0.8$. We clearly see that the greater $\delta$ the lower is $C$. We have set $\Sigma = 1$.}
\label{fig1a}
\end{figure}

\begin{figure}[ht]
\centerline{\psfig{file=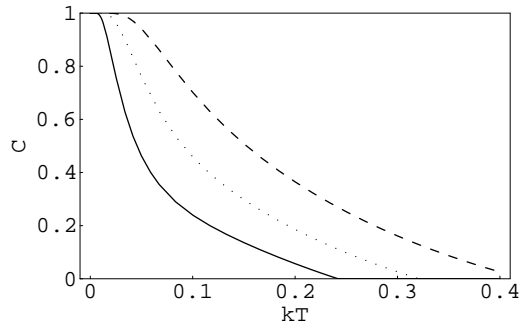, width=7cm}}
\vspace*{8pt}
\caption{The dependence of the concurrence $C$ for the XY model with the absolute temperature $kT$. The solid line represents $\delta = 1.2$, the dotted line shows $\delta = 1.4$, and the dashed line is for  $\delta = 1.7$. We clearly see now that the greater $\delta$ the \textit{greater} is $C$. We have set $\Sigma = 1$.}
\label{fig1b}
\end{figure}

 Setting $J_{z} = 0$ in Eq.~(\ref{geralxyz}) we obtain the following expression for the concurrence of the anisotropic XY model:
\begin{eqnarray}
C = \mbox{max}\left\{ 0, \frac{\sinh\left(|\gamma|\right)-\cosh\left(\beta\right)}{\cosh\left(\gamma\right)+\cosh\left(\beta\right)} \right\}, &  \mbox{if} \; |\delta| < 1,  \label{xyaniso1} \\
C = \mbox{max}\left\{ 0, \frac{\sinh(|\beta|)-\cosh(\gamma)}{\cosh(\gamma)+\cosh(\beta)} \right\}, & \mbox{if} \; |\delta| \ge 1. \label{xyaniso2}
\end{eqnarray}
Looking at Eqs.~(\ref{xyaniso1}) and (\ref{xyaniso2}) we see two regions of anisotropy. The first region, $|\delta| < 1$, was studied by Wang \cite{wang} and by Kamta and Starace \cite{kamta}. They have shown that increasing the anisotropy parameter $|\delta|$ the concurrence decreases for a given temperature T and that when $|\delta| = 1$ the concurrence is zero for all T (Ising model). However, in the second region, $|\delta| > 1$, we see that the concurrence \textit{increases} as we increase the anisotropy parameter $\delta$ and that the critical temperature $T_{c}$ increases as we increase the anisotropy. We also see that in both regions the concurrence is a monotonically decreasing function of the temperature. Figs.~(\ref{fig1a}) and (\ref{fig1b}) show this behavior.

It is interesting to note that the conditions on $|\delta|$ given in Eqs.~(\ref{xyaniso1}) and (\ref{xyaniso2}) are equivalent to $J_{x}J_{y} > 0$ and $J_{x}J_{y} \le 0$ respectively. This implies that the anisotropy increases the concurrence if $J_{x}$ and $J_{y}$ have different signs.

\section{XXX Model}

\begin{figure}[ht]
\centerline{\psfig{file=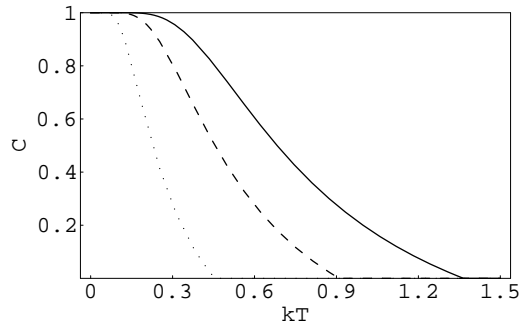, width=7cm}}
\vspace*{8pt}
\caption{The dependence of the concurrence $C$ with the absolute temperature $kT$ for three values of the coupling constant $J$ for the XXX model. The solid line represents $J= 1.5$, the dashed line shows $J= 1$, and the dotted line is for  $J = 0.5$.}
\label{fig3}
\end{figure}

When $J_{x}=J_{y}=J_{z}=J$ we have the XXX model \cite{vedral}, which implies that $\beta = 0$ and $\gamma = 2 \alpha$. Eq.~(\ref{geralxyz}) then shows that $C = 0$ if $J \le 0$ and
\begin{equation}
C  =  \mbox{max}\left\{ 0, \, \frac{1-3e^{-4\alpha}}{1+3e^{-4\alpha}}  \right\}, \; \mbox{if} \; J > 0,  \label{xxx1}
\end{equation}
It is interesting to note that contrary to the XY isotropic model \cite{wang} the concurrence for the ferromagnetic XXX model is always zero \cite{vedral}. This can be understood if we consider the eigenvectors of the system. For $J<0$ we have a degeneracy for the ground state, which is formed by the triplets. Therefore $\rho(T =0)$ $=$ $(1/3)\left( \left| \Psi^{+} \right>\left< \Psi^{+} \right| +\left| \Phi^{+} \right>\left< \Phi^{+} \right| + \left| \Phi^{-} \right>\left< \Phi^{-} \right| \right)$, which is a non-entangled state. And increasing the temperature the singlet mixes with the triplets in a non-entangled state. However, when $J>0$, the ground state is the singlet, a maximally entangled state, and we have $C =1$. Increasing the temperature reduces the entanglement because we mix the triplets with the singlet. See Fig.~(\ref{fig3}).

\section{XXZ Model}
\begin{figure}[ht]
\centerline{\psfig{file=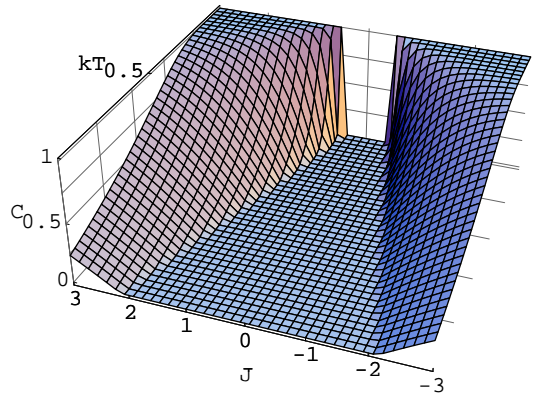, width=7cm}}
\vspace*{8pt}
\caption{The concurrence $C$ as a function of the absolute temperature $kT$ and of $J$. We have set $J_{z}$ $=$ $-0.5$. It is clear that there exists a region where $C$ $=$ $0$ for any $T$.}
\label{fig4}
\end{figure}

 If $J_{z}\ne J$ and $J_{x}=J_{y}=J$ we have the XXZ model. Now $\beta = 0$ and Eq.~(\ref{geralxyz}) gives $C = 0$ if $2\alpha \le -|\gamma|$ and
\begin{equation}
C = \mbox{max} \left\{ 0, \; \frac{e^{2\alpha}\sinh(|\gamma|)-1}{e^{2\alpha}\cosh(\gamma)+1} \right\}, \; \mbox{if} \; 2\alpha > -|\gamma|.
\end{equation} 
Again we can understand why for  $2\alpha$ $\le$ $-|\gamma|$ there exists no entanglement even at $T$ $=$ $0$ by looking at the ground state of $\rho$. In this region $J_{z}$ $<$ $0$ and $\left| \Phi^{\pm} \right>$ are the degenerated ground states. Therefore, $\rho(T=0)$ $=$ $(1/2)$ $\left( \left| \Phi^{+} \right> \left< \Phi^{+} \right|\right.$ $+$ $\left.\left| \Phi^{-} \right> \left< \Phi^{-} \right|  \right)$ $=$ $(1/2)$ $\left( \left|00\right>\left<00\right|\right.$ $+$ $\left.\left|11\right>\left<11\right| \right)$, which is a separable state. By increasing $T$ we mix $\left| \Psi^{\pm} \right>$ with $\left| \Phi^{\pm} \right>$, producing a non-entangled state. On the other hand, when  $2\alpha$ $>$ $-|\gamma|$ the ground state is the singlet, a maximally entangled state. By increasing the temperature the components of the triplet mix with the singlet decreasing the concurrence. See Fig.~(\ref{fig4}).

It is worthy of mention that we have numerically observed that when $J_{z} > 0$, an increase of its value increases the entanglement for a given $T$. We also note that whatever the sign of $J$, increasing its modulo always increases entanglement. These two behaviors can be understood analyzing the eigenvalues of the system: increasing the absolute value of $J$ or increasing $J_{z}$ increases the proportion of singlet ($J$ positive) or the proportion of $\left| \Psi^{+}\right>$ ($J$ negative) in the thermal state. These two facts are responsible for an increment in the concurrence.
  
\section{XYZ Thermal State: Detailed Study}

\begin{figure}[ht]
\centerline{\psfig{file=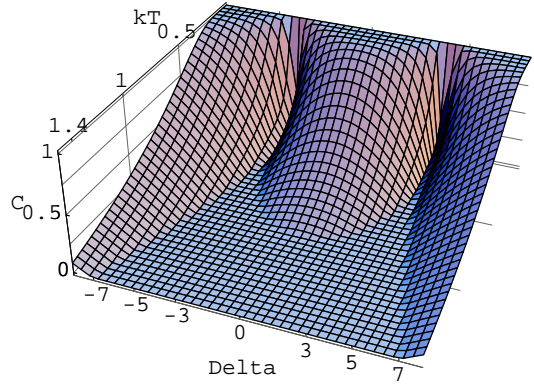, width=7cm}}
\vspace*{8pt}
\caption{The concurrence $C$ as a function of $\Delta$ and of $kT$. We see that there are regions where an increase in the anisotropy \textit{increases} $C$ and that $C$ is a monotonically decreasing function of $kT$. $\Sigma = 2$ and $J_{z} = 1$.}
\label{fig5aa}
\end{figure}

\begin{figure}[ht]
\centerline{\psfig{file=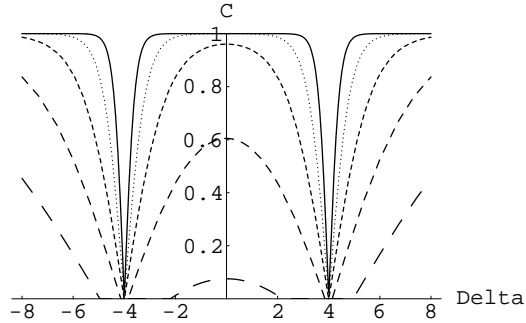, width=7cm}}
\vspace*{8pt}
\caption{The concurrence $C$ as a function of $\Delta$ for various values of $kT$. For the solid line $kT = 0.05$, dotted line $kT=0.1$, short dashed $kT=0.2$, dashed $kT=0.4$, and long dashed $kT=0.8$. We clearly see that for $|\Delta| > 4$, i. e. ($2\alpha < |\beta|-|\gamma|$), the higher the anisotropy (greater $|\Delta|$) the more entangled is the system. $\Sigma = 2$ and $J_{z} = 1$.}
\label{fig5bb}
\end{figure}

We now analyze the XYZ model. $J_{x}, J_{y}$, and $J_{z}$ are allowed to assume any values and we must study Eq.~(\ref{geralxyz}) in its general form. We first observe an interesting situation. Whenever $2\alpha = |\beta| - |\gamma|$ we have null concurrence even for $T=0$. This condition is equivalent to $2J_{z}=|\Delta| - |\Sigma|$. This implies that systems with coupling constants near this region are not useful in generating entanglement. Figs.~(\ref{fig5aa}, \ref{fig5bb})  and (\ref{fig6aa}, \ref{fig6bb}) highlight this behavior. 
   
Eq.~(\ref{geralxyz}) implies that for fixed $J_{z}$ there exist regions where increasing the anisotropy parameter $\delta$ we \textit{increase} the concurrence. In the region where $2\alpha < |\beta|-|\gamma|$ the higher the anisotropy the more the system is entangled. However, in the region where  $2\alpha > |\beta|-|\gamma|$, which reduces to the region studied by Refs.~\refcite{kamta} and \refcite{wang} if we put $J_{z} = 0$, the anisotropy decreases entanglement. Figs.~(\ref{fig5aa}) and (\ref{fig5bb}) illustrate this.

\begin{figure}[ht]
\centerline{\psfig{file=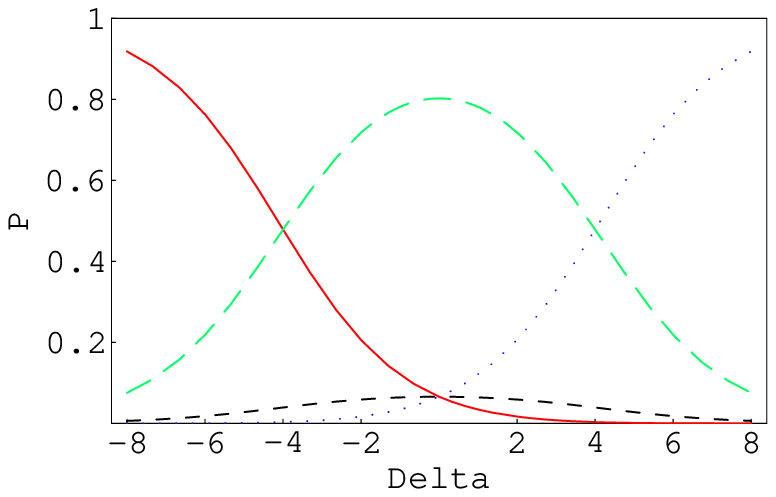, width=7cm}}
\centerline{\psfig{file=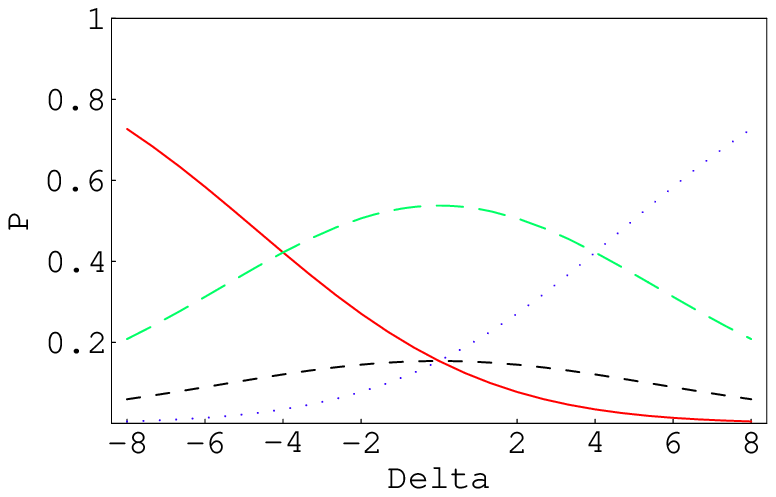, width=7cm}}
\vspace*{8pt}
\caption{The probability distribution $P$ of the eigenvectors of the XYZ Hamiltonian in the thermal state as a function of $\Delta$. The solid/red line gives $P$ for $\left|\Phi^{+}\right>$, dotted/blue for $\left|\Phi^{-}\right>$ , short dashed/black for  $\left|\Psi^{+}\right>$, and long dashed/green for $\left|\Psi^{-}\right>$. For large $|\Delta|$ and $|\Delta| \approx 0$ only one maximally entangled state dominates, justifying why we have high concurrences in this region. And more, as we increase the temperature, we see that near $\Delta = 0$ the triplets mix with the singlet decreasing  $C$. For large $|\Delta|$ this mixing only occurs for higher temperatures. $J_{z}=1$ and $\Sigma = 2$. Top: $kT = 0.4$. Bottom: $kT = 0.8$.}
\label{fig5a}
\end{figure}

\begin{figure}[ht]
\centerline{\psfig{file=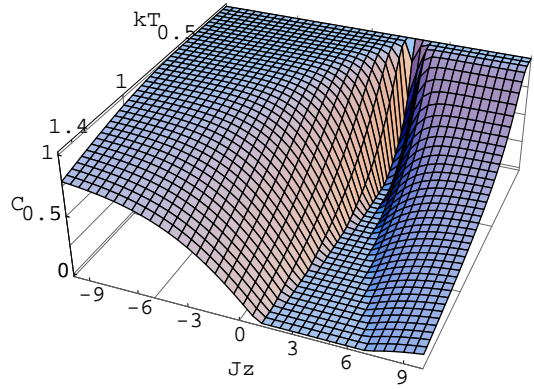, width=7cm}}
\vspace*{8pt}
\caption{The dependence of the concurrence $C$ as a function of $J_{z}$ and $kT$. As we move away from $J_{z} = 3$, i. e. ($|\Delta| - |\Sigma|)/2$), we get higher values for $C$, which is a decreasing function of $kT$. $\Delta = 7$ and $\Sigma = 1$.}
\label{fig6aa}
\end{figure}

\begin{figure}[ht]
\centerline{\psfig{file=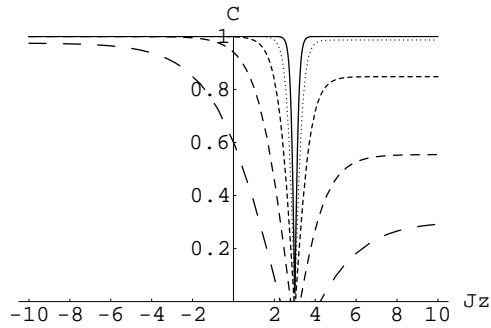, width=7cm}}
\vspace*{8pt}
\caption{ The concurrence $C$ is a function of $J_{z}$ for different values of $kT$. For the solid line $kT = 0.05$, dotted line $kT=0.1$, short dashed $kT=0.2$, dashed $kT=0.4$, and long dashed $kT=0.8$.  In the region where $2\alpha > |\beta|-|\gamma|$ increasing $J_{z}$ we get $C\approx 1$ only for $kT\approx 0$. $\Delta = 7$ and $\Sigma = 1$.}
\label{fig6bb}
\end{figure}

We can understand the behavior of $C$ as we vary $\Delta$ looking at the probability distribution $P$ of the four eigenvectors of the XYZ Hamiltonian in the thermal state. Here, 
\begin{eqnarray}
P_{\Phi^{\pm}} = Tr\left\{\left|\Phi^{\pm}\right>\left<\Phi^{\pm}\right|\rho\right\} = \frac{\exp{\left(-\lambda_{\Phi^{\pm}}/kT\right)}}{Z},  \label{prob1}\\
P_{\Psi^{\pm}} = Tr\left\{\left|\Psi^{\pm}\right>\left<\Psi^{\pm}\right|\rho\right\} = \frac{\exp{\left(-\lambda_{\Psi^{\pm}}/kT\right)}}{Z}. \label{prob2}
\label{extra1}
\end{eqnarray}  
We see that when $2\alpha = |\beta|-|\gamma|$ the thermal state is an equal mixture of $\left| \Phi^{\pm}  \right>$  ($\Phi^{-}$ for $\beta > 0$ and $\Phi^{+}$ for $\beta <0$) and $\left|\Psi^{-} \right>$ plus a tiny fraction of $\left|\Psi^{+} \right>$. The density matrix that describes the system can be written as $\rho = (1/2-\epsilon/2)\left(\left| \Phi^{\pm}  \right>\left< \Phi^{\pm}  \right|+ \left| \Psi^{-}  \right>\left< \Psi^{-}  \right|\right) + \epsilon \left| \Psi^{+}  \right>\left< \Psi^{+}  \right|$, where $\epsilon \ll 1$. This density matrix is separable for $\epsilon \leq 1/2$, which explains why in the region where $2\alpha = |\beta|-|\gamma|$ we have no entanglement. See Fig.~(\ref{fig5a}).

We can also understand why for $2\alpha < |\beta|-|\gamma|$ the anisotropy increases entanglement studying Eq.~(\ref{prob1}) and (\ref{prob2}). In this region, we see that an increase in the anisotropy (higher $\Delta$) produces a thermal state almost dominated by only one maximally entangled state, which implies an increase in $C$. On the other hand, if we are in the region where $2\alpha > |\beta|-|\gamma|$, an increase in the anisotropy produces a mixed state of two maximally entangled state, which causes a decrease in $C$. In the limit where $2\alpha = |\beta|-|\gamma|$ we have a mixture of these two states in equal proportions, implying $C=0$. See Fig.~(\ref{fig5a}).

Fixing $J_{x}$, $J_{y}$, and varying $J_{z}$ we see that the concurrence increases if we pick values of $J_{z}$ greater or lower than $(|\Delta| - |\Sigma|)/2$. There exists, however, a value of $J_{z}$ beyond which $C$ does not increase anymore. This behavior is more drastic if we are in the region where  $2\alpha > |\beta|-|\gamma|$. There, only for $kT\approx 0$ we obtain $C\approx 1$. For any other value of $kT$, increasing $J_{z}$ makes $C \rightarrow C_{max}$, where $C_{max} < 1$. And more, the higher $kT$ the lower is the value of $C_{max}$. If we are in the region where $2\alpha < |\beta|-|\gamma|$, decreasing $J_{z}$ we still can asymptotically obtain $C = 1$ for $kT > 0$. See Figs.~(\ref{fig6aa}) and (\ref{fig6bb}). 

\begin{figure}[ht]
\centerline{\epsfig{file=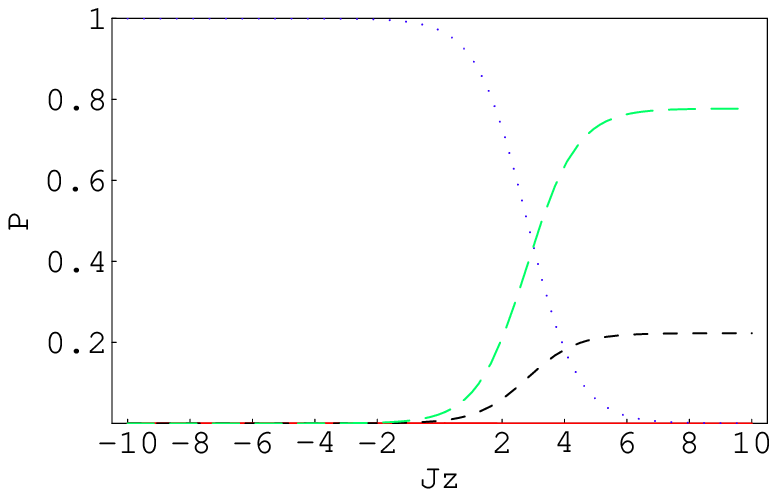, width=7cm}}
\centerline{\epsfig{file=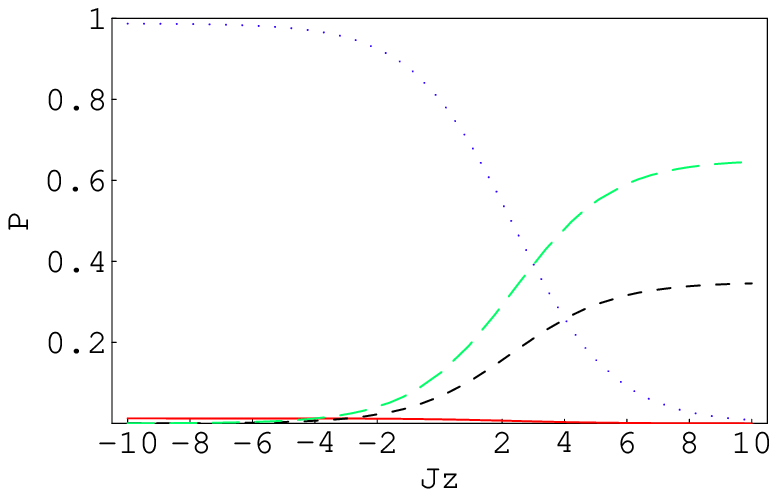, width=7cm}}
\caption{The probability distribution $P$ of the eigenvectors of the XYZ Hamiltonian in the thermal state as a function of $J_{z}$. The solid/red line gives $P$ for $\left|\Phi^{+}\right>$, dotted/blue for $\left|\Phi^{-}\right>$ , short dashed/black for  $\left|\Psi^{+}\right>$, and long dashed/green for $\left|\Psi^{-}\right>$. As we move away from $J_{z} = 3$ one of the four maximally entangled states begins to dominate, explaining why $C$ increases. $\Delta=7$ and $\Sigma = 1$. Top: $kT = 0.4$. Bottom: $kT = 0.8$.}
\label{fig6a}
\end{figure}

We can again get a physical picture of the behavior of $C$ as a function of $J_{z}$ studying the probability distribution $P$ of the four eigenvectors of the XYZ Hamiltonian in the thermal state. We see that as we move away from $J_{z} = (|\Delta| - |\Sigma|)/2$ one of the four maximally entangled states begins to dominate, explaining why $C$ increases. But only in the region where  $2\alpha < |\beta|-|\gamma|$ there exists for $kT > 0$ a value of $|J_{z}|$ beyond which the probability distribution is zero for the other three Bell states, implying $C=1$. If we are in the region where $2\alpha > |\beta|-|\gamma|$ we get for large $J_{z}$ a reasonable contribution  of another maximally entangled state $(\left| \Psi^{+}\right>)$, justifying why $C < 1$ in this region. See Fig.~(\ref{fig6a}).   

We end our study of the XYZ chain pointing that we have numerically searched for a set of coupling constants that should give $\partial C/\partial (kT) > 0$. If such set existed we would have found a region where increasing the temperature causes an increase in the entanglement with no need of an external magnetic field \cite{nielsen,vedral,kamta}. We tested the sign of $\partial C/\partial (kT)$ for all combinations of $J_{x}, J_{y}$, and $J_{z}$ ranging from $-2$ to $2$ in increments of $0.01$ and from $-50$ to $50$ in increments of $0.5$. We have not found any set of constants where $\partial C/\partial (kT) > 0$. This result suggests that we must have an external magnetic field in order to get a positive derivative.   
 
\section{Conclusion}

In this article, we have studied the thermal entanglement for the XYZ Heisenberg chain. We restricted our attention to chains of two qubits with no external magnetic field. 

We reviewed the well known results for the XY and XXX models and analyzed the XXZ and XYZ models in detail. 

We presented a general analytical expression for the concurrence of the XYZ model and then particularized to some simple cases. We have shown that there exist regimes in the XY and XYZ models where an increase of the anisotropy in the coupling constants causes an increase in the amount of entanglement for a given temperature $T$. In these regions, the critical temperature, beyond which the entanglement is zero, also increases with the anisotropy. This result is very interesting since up to now only regions where a decrease of the entanglement with anisotropy were studied.

We have numerically checked that there does not exist any combination of the coupling constants where the entanglement increases with $T$. This result suggests that we must have an external magnetic field applied to the qubits to prevent the entanglement to monotonically decrease with $T$, i. e., we need external fields to obtain a region where the entanglement increases as we increase T.       

Finally, we want to mention that we are just beginning to understand the relation between entanglement and temperature in Heisenberg chains. Here we have dealt with only two qubits and nearest neighbors interactions. It would be interesting to study chains with a large number of qubits as well as chains with long range interactions.
 
\section*{Acknowledgments}
We would like to express our gratitude to the funding of Funda\c{c}\~ao de Amparo \`a Pesquisa do Estado de S\~ao Paulo (FAPESP). We thank L. F. Santos and C. O. Escobar for a careful reading of the manuscript.

\end{document}